\documentclass[11pt]{article}
\usepackage{amsfonts}
\usepackage{amssymb}
\usepackage{amsthm}
\usepackage{amsmath}
\usepackage{newlfont}
\usepackage{newlfont}
\usepackage{bm}
\usepackage{txfonts}
\usepackage{graphics}
\usepackage{graphicx}
\usepackage{color}
\usepackage{indentfirst}
\begin{document}

\title{{\bf Edge dislocation core structures in FCC metals determined from ab initio calculations combined with the improved Peierls-Nabarro equation} \footnote{Project supported
by the National Natural Science Foundation of China (11074313) and Project No. CDJXS11102211 supported by the Fundamental
Research Funds for the Central Universities.}}

\author{  Rui Wang\footnote{Corresponding author.\ E-mail: rcwang@cqu.edu.cn;  tel: $+$86 13527528737},\ \ Shaofeng Wang,\ \ Xiaozhi  Wu \\
{\small  {  Institute for Structure and Function and Department of Physics, Chongqing
University, Chongqing 400044, P. R. China }} }
\date{}
\maketitle \vskip 0.25in \noindent {\bf Abstract} {\small  We have
employed the improved  Peierls-Nabarro (P-N) equation to study the
properties of 1/2$\langle110\rangle$ edge dislocation in \{111\} plane
in FCC metals Al, Cu, Ir, Pd, and Pt. The generalized-stacking-fault
 energy (GSFE) surface entering the equation is calculated by using
first-principles density functional theory (DFT). The accuracy of
the method has been tested by calculating values for various
stacking fault energies which favorably compare with the previous
theoretical and experimental results. The core structures, including
the core widths both of the edge and screw components, dissociation
behavior for edge dislocations have been investigated. The
dissociated distance between two partials for Al in our calculation
agrees well with the values obtained from the numeric simulation with DFT
and molecular dynamics simulation, as well as experiment. Our
calculations show that it is preferred to create partial dislocation
in Cu, and to be easily observed full dislocation in Al, Ir, Pd, and
especially Pt. }

\vskip 0.15in
 \noindent {\bf PACS:} 61.72.Bb; 71.15.Mb.

 \noindent {\bf Keywords:}  {\small
FCC Metals;  Ab initio calculations;  Generalized-stacking-fault
energy (GSFE);  Dislocation properties; Improved Peierls-Nabarro equation.}

\vskip 0.5in \baselineskip 20pt
\section{Introduction}
It is widely accepted that the dislocations have a direct influence
on the deformation mechanism of materials \cite{Hirth}, and there
has been a great deal of interest in describing accurately the
dislocation core structure on an atomic scale because of its
important role in many phenomena of crystal plasticity
\cite{Duesbery,Vitek}. In FCC metals dislocations can reduce elastic
energy by dissociating into Shockley partial dislocations connected
by a stacking fault \cite{Hirth}. But as the core regions of two partials overlap, whether the partials can be observed has not been governed by the stacking fault energy but also generally by atomic interactions.   The dislocations in
Al have always been focused on in the past twenty years
\cite{Woodward,Hollerbauer, Lu1, Lu2, SchoeckAl, Wang, Olmsted,
Srinivasan,Mishin,Hartford}. The accurate prediction of the
atomic-scale core geometries for dislocations in Al is needed to
inform higher-length scale models of plasticity in Al alloys, as
well as as other FCC metals \cite{Woodward}. In addition, it is well
known that the nature of slip cannot be described in terms of an
absolute value of stacking fault energy and a correct interpretation
requires the generalized-stacking-fault energy (GSFE) curve,
involving both intrinsic (stable) and unstable stacking fault
energies \cite{Swygenhoven}.

There are usually two types of theoretical approaches which have
been employed to study the core properties of dislocations. The
first method is based on direct atomistic simulations or
first-principles calculations, and the second method is based on the
the framework (Peierls-Nabarro) P-N model. Empirical interatomic
potentials involve the fitting of parameters to a predetermined
database and hence may not reliable in describing the core
properties. On the other hand, first-principles calculations, though
considerably more accurate, are computationally expensive for
studies of dislocation properties. The P-N model, in which the
discreteness of crystal is neglected, is essentially a continuum
treatment, but the dislocation core, the region of inelastic
displacement, is given an approximate atomistic description. In
addition, the partials in FCC metals are mixed dislocations with
both edge and screw components \cite{Woodward}.  Schoeck and Mryasov et al adopted the generalized
2D P-N model to consideration the two-component displacement field
of dissociated dislocation \cite{SchoeckAl,Schoeck2005,Mryasov}, but
discreteness of crystals is neglected. Recently, the 2D improved P-N
dislocation equation taken into account the discreteness of crystals
based on the lattice dynamics and the symmetry principle has been
obtained \cite{wang2002,wang2009}. Both for the P-N model and
improved P-N equation \cite{wang2009}, the forces in the dislocation
core, where the atomic atomic-scale discreteness really counts, are
currently approximated with the GSFE surface ($\gamma-$surface)
\cite{Vitek,Hartford,Vitek1,Vitek2}. The GSFE surface is the interplanar
potential energy, which can currently be calculated accurately from
ab initio calculations,  for sliding one half of a crystal over the
other half.

In this paper, we carry out first principles calculations on the
GSFE surfaces ($\gamma -$surface) for FCC metals Cu, Al, Ir, Pd, and
Pt. The $\gamma -$surfaces are applied in the improved P-N equation
to the edge dislocations. The accuracy of the method has been tested
by calculating values for various stacking fault energies which
favorably compare with the previous theoretical and experimental
results. Dislocation structures are calculated, including the core
widthes and dissociated distance between two partials.  This paper
is organized as follows. In section 2, we carry out the first
principles calculation of GSFE surface on (111) plane in FCC metals
Cu, Al, Ir, Pd, and Pt. In section 3, the core structure of
dissociated dislocation is obtained from the improved P-N equation
with variational method, and dislocation properties have  also been
discussed.

\section{ First principles calculations of
generalized-stacking-fault-energy surface}
 \noindent{\large \bf 2.1.
Computational Methodology}

In the present study total-energy calculations based on the density
functional theory (DFT) embodied in the Vienna ab simulation package
(VASP) \cite{Kresse1, Kresse2, Kresse3} is employed. The
Perdew-Burke-Ernzerhof (PBE) \cite{Perdew1,Perdew2}
exchange-correlation functional for the
generalized-gradient-approximation(GGA) is used. A plane-wave basis
set is employed within the framework of the projector augmented wave
(PAW) method \cite{Blochl, Kresse4}. On the basis of our tests, we
have chosen the plane-wave energy cutoff of 500 eV for all
calculated metals. An initial calculation is undertaken to determine
the optimum lattice constants as well as the elastic constants.  For
the first-Brillouin-zone integrals, reciprocal space is represented
by Monkhorst-Pack-special k-point scheme \cite{Monkhorst} with
$15\times15\times15$ grid meshes for these initial calculations,
while the generalized-stacking-fault  energy (GSFE) calculations
employ $21\times21\times3$ grid meshes for our calculated FCC metals
Cu, Al, Ir, Pd, and Pt. The equilibrium theoretical lattice
structures are determined by minimizing the Hellmann-Feynman force
on the atoms and the stress on the unit cell. The convergence of
energy and force are set to $1.0\times10^{-6}eV$ and
$1.0\times10^{-4}eV/{\AA}$, respectively. In Table 1, we give the
all equilibrium lattice constants and elastic modulus for all
studied FCC metals in our calculation, and it is shown that the
results agree well with the experiment values
\cite{Handbook1,Handbook2}.

 \noindent{\large \bf 2.1.
The generalized-stacking-fault energy surfaces}

In the present study, we calculate the GSFE surfaces in the
closed-packed (111) surface which is the trigonal lattice, since the
slip between the closed-packed surfaces is most easily for FFC
metals. The ideal FCC structures have the configuration $\cdots
ABCABC \cdots$ stacking sequence of the atomic planes. To simulate
the block shearing process we use a slab consisting of twelve atomic
layers in the $\langle111\rangle$ direction. Between periodically
repeated slab the vacuum space of 15${\AA}$ normal to (111) plane is
chosen to avoid interactions between two slabs. The stacking-fault
energies (SFEs) for the slab consisting of twelve and fourteen
atomic layers are reasonably close. In addition, we find that the
fluctuations of calculated results for vacuum gap of 15${\AA}$ and
18${\AA}$ are less than $1.0\%$. So, the adequate convergence with
respect to the supercell size consisting of twelve atomic-layers
thick slab and vacuum gap of 15${\AA}$ is indicated. The slab
calculations is appropriate since the fault interactions are short
ranged \cite{Hartford}. Comparison with a six atomic-layers thick
slab calculations of intrinsic-stacking-fault (ISF) by Hartford et
al \cite{Hartford}, our twelve atomic-layers thick slab calculation
is more accurate, but with great computational consumption.

The SFEs  are determined as a difference of total energies for two
sides of the slab, in which there are six atomic layers for below
and above, designed to simulate faults with vectors $\mathbf{u}=0$
and $\mathbf{u}\neq0$. The stacking-fault (SF) vector $\mathbf{u}$,
which is a two-dimension (2D) vector $(u_{x}, u_{y})$, is the
displacement of the upper half-crystal relative to the one below.
The ISF configuration corresponds to a
slip of $\frac{a}{\sqrt{6}}$ in the $\langle112\rangle$ direction,
resulting in the SF vector
$\mathbf{u}=\frac{a}{\sqrt{6}}(\frac{\sqrt{3}}{2}, \frac{1}{2})$
(where $a$ is the lattice constant). The unstable stacking-fault
energy (USFE) $\gamma_{\mathrm USF}$ corresponds to the lowest
energy barrier that needs to be crossed for the slip from the ideal
configuration to the ISF configuration in the $\langle112\rangle$
direction. In order to obtain the correct SFEs, structural
optimizations must be considered in this study. The relaxations
perpendicular to the (111) plane are carried out by a combined
optimization of supercell volume and atomic coordinates in shifted
configuration.

To obtain the GSFE surfaces of FCC metals,  we have calculated SFEs
for the slips along the $\langle112\rangle$ and $\langle110\rangle$
directions, respectively. Taking Al as an example,  the fitted GSFE
surface is shown in Fig. 1, while the GSFE surfaces for other FCC
metals has the similar shapes.  The first energy maximum encountered
along the $\langle112\rangle$ direction is the USFE
$\gamma_{\mathrm{USF}}$, which represents the lowest energy barrier
for dislocation nucleation\cite{Rice}; the first energy minimum
value at $1/6 \langle112\rangle$ $(a/\sqrt{6})$ corresponds to the
ISF configuration, where a full dislocation dissociates into a pair
of Shockley partials. The projection of the GSFE surface on the
$\langle110\rangle$ direction are symmetric with respect to the slip
displacement of $\sqrt{2}a/4$. In Fig. 2, we show the projection of
the GSF energy surface on the $\langle112\rangle$ (Fig. 3(a)) and
$\langle110\rangle$ (Fig. 3(b)) directions. The values of unstable
stacking-fault (SF) energy along the $\langle110\rangle$ direction
are found to be larger than one along the $\langle112\rangle$
direction. The energy values of the various stacking faults obtained
from the DFT calculations are summarized in Table 2. We also compare
our results with the previous theoretical results
\cite{Hartford,Zhang1,Wei} and experimental results
\cite{Hirth,Grujicic,Thorntor}. The calculated results of ISF
energies (ISFEs) $\gamma_{\mathrm{ISF}}$ are in good agreement with
the experimental values. Comparison with the results obtained from
the EAM (embedded atom method) \cite{Zhang1} or MEAM (modified
EAM)\cite{Wei} by Zhang et al, our results agree better with the
experimental results. Our results for Al and Pd are also in
consistent with the results obtained from Hartford et al
\cite{Hartford} by first-principles calculations. The value of
$\gamma_{I}$ for Pt is 282$\mathrm{mJ/m^{2}}$ which is in better
agreement with the experimental value 322$\mathrm{mJ/m^{2}}$ than
the results 111$\mathrm{mJ/m^{2}}$ obtained from MEAM \cite{Wei}.

The GSFE surface $\gamma(\mathbf{u})$ satisfies the translational
symmetry,
\begin{equation}
\gamma (\mathbf{R}+\mathbf{u})=\gamma (\mathbf{u}),
\end{equation}
where $\mathbf{R}$ is the lattice vector.  The GSFE surface can be
represented with the aid of reciprocal lattice by a 2D Fourier
series which reflects the full symmetry of the (111) plane,
\begin{equation}\label{gsf}
\gamma (\mathbf{u})=\sum _{\mathbf{G}}\gamma _{\mathbf{G}}
e^{i\mathbf{G} \cdot \mathbf{u}}
\end{equation}
where $\mathbf{G}$ is the reciprocal lattice vector.  While the
nearest neighbor approximation in reciprocal lattice space is taken
into account
\begin{eqnarray}\label{gamma}
\gamma (u^{x},u^{y})&=&\gamma _{0} + \gamma _{1}\Big\{\cos
\Big(\frac{4\pi u^{y}}{\sqrt{3}a}\Big)+\cos \Big(\frac{2\pi
u^{x}}{a}+ \frac{2\pi u^{y}}{\sqrt{3}a}\Big)+\cos \Big(\frac{2\pi
u^{x}}{a}-
\frac{2\pi u^{y}}{\sqrt{3}a}\Big)\Big\} \nonumber \\
& &+\gamma _{2}\Big\{\sin \Big(\frac{4\pi u^{y}}{\sqrt{3}a}\Big)-
\sin\Big(\frac{2\pi u^{x}}{a}+\frac{2\pi u^{y}}{\sqrt{3}a}\Big)
+\sin\Big(\frac{2\pi u^{x}}{a}-\frac{2\pi
u^{y}}{\sqrt{3}a}\Big)\Big\}
\end{eqnarray}
The coefficients $\gamma _{0}$, $\gamma _{1}$ and $\gamma _{2}$ can
be determined by the fitting to the  GSFE surface that can be
obtained by our first-principles  calculations. $\gamma _{0}$
represents the ground state energy , which does not influence the
shape of GSFE surface and can be chosen to zero, so only $\gamma
_{1}$ and $\gamma _{2}$ need to be fixed.  We also list the fitted
parameters $\gamma _{1}$ and $\gamma _{2}$ of Eq. (\ref{gamma}) for
all calculated FCC metals in Table 2. Eq. (\ref{gamma}) is accurate
to describe translational and reflection symmetry of the GSFE
surface on (111) plane in FCC metals, and facilitates the
computation of dislocation properties.

\section{The dissociation of dislocation and the governing equation}

\noindent{\large \bf 3.1. The improved Peierls-Nabarro dislocation
equation}

 In the (111) plane of FCC metals, the primary slip
system is $\langle112\rangle$(111), so dislocations with Burgers'
vector $ \frac{1}{2}\langle 110 \rangle$ will
 energetically favorable dissociate into two Shockley partials
 connected by a stacking-fault\ (SF) ribbon according to Frank's rule \cite{Hirth}.
The dissociation process is described by the following reaction
\begin{equation}
\frac{1}{2}[\overline{1}10] \rightarrow \frac{1}{6}[
\overline{1}12]+SF+\frac{1}{6}[ \overline{1}1\overline{2}].
\end{equation}
The dissociation can be understood from the GSFE surfaces for slip
plane (111). Fig.2 shows that the lowest energy path happens to be
along the $\langle112\rangle$ direction, and perfect dislocations
(Burgers' vector $\mathbf{b}=\frac{1}{2}[\overline{1}10]$) is likely
to dissociate into partial dislocations (Burgers' vectors are
$\mathbf{b}_{1}=\frac{1}{6}[ \overline{1}12]$ and
$\mathbf{b}_{2}=\frac{1}{6}[ \overline{1}1\overline{2}]$) due to the
lower energy barrier. The two partial dislocations are $60^{\circ}$
fractional dislocations viz. the included angle between the
dislocation line and the Burgers vector is $60^{\circ}$. The
magnitude of the total Burgers vector is $b=\sqrt{a}/2$ and $a$ is
the lattice constant for FCC metal. The Burgers vectors of the two
partial dislocations have magnitudes $b_{1}=b_{2}=\sqrt{3}b/3$.
Hereafter, let the dislocation lines parallel to the y-axis
($[11\bar{2}]$ direction) and perpendicular to the total Burgers
vector of edge dislocation ($[\bar{1}10]$ direction). The
x-component and y-component respectively represent the edge
component and screw component of partial dislocations for
dissociation from edge dislocation.

The 2D dislocation equation for straight dislocations that describe
the balance of atoms on the border based on the lattice dynamics and
the symmetry principle takes the following form \cite{wang2009}
\begin{eqnarray}\label{mpn}
-\frac{\beta_{e}}{2}
\frac{d^{2}u^{x}}{dx^{2}}-\frac{K_{e}}{2\pi}\int_{-\infty}^{+\infty}\frac{dx^{'}}{x^{'}-x}
\Big(\frac{du^{x}}{dx}\Big)\Big|_{x=x^{'}}&=&f^{x}(u^{x},u^{y}), \nonumber\\
 -\frac{\beta_{s}}{2}
\frac{d^{2}u^{y}}{dx^{2}}-\frac{K_{s}}{2\pi}\int_{-\infty}^{+\infty}\frac{dx^{'}}{x^{'}-x}
\Big(\frac{du^{y}}{dx}\Big)\Big|_{x=x^{'}}&=&f^{y}(u^{x},u^{y}),
\end{eqnarray}
where $K_{e}$ and $K_{s}$ are the energy factors of the edge and
screw dislocations, $u^{x}$ and $u^{y}$ are the edge and screw
components of displacements. For the isotropic solid, $K_{e}=\mu
\sigma /(1-\nu)$ and $K_{s}=\mu \sigma$, with $\mu$ being shear
modulus and $\nu$ Poisson's ratio, $\sigma$ is the area of the
primitive cell of the misfit plane. The shear modulus and the
Poisson's ratio can be expressed as the second-order elastic
constants $c_{ij}$, i.e., $\mu=(c_{11}-c_{12})/2$ and
$\nu=c_{12}/(c_{11}+c_{12})$ (the values as shown in Table 1).  The
coefficients of the second-order derivations $\beta_{e}$ and
$\beta_{s}$ relate with the acoustic phonon velocity and the lattice
geometry structure
\begin{eqnarray}
\beta_{e}& =&\frac{3}{4}\Omega \mu \Big(\frac{2-2\nu}{1-2\nu}-\tan ^{2}\theta\cos ^{2}\phi\Big),\\
\beta_{s}&=&\frac{3}{4}\Omega \mu (1-\tan ^{2}\theta\sin ^{2}\phi),
\end{eqnarray}
where $\theta$ and $\phi$ are the orientation angles of the relative
position of a pair of neighbor atoms in the intrinsic frame with the
axes given by the polarization directions, and $\theta =\pi /4$ and
$\phi =\pi /6$ for FCC crystals, $\Omega$ is the volume of the
primitive cell. Comparing with the generalized 2D P-N equation in
P-N model, there are two extra second-order derivations that
represent the discreteness effect of crystal in the new equation. From exactly solvable models \cite{wang2002, wang2009} in which  discreteness of a lattice can be fully considered, one clearly sees the discreteness correction appearing in terms of the second-differential of the displacement.
 Physically, the integro term in Eq. (\ref{mpn}) represents a long range interaction which is inversely proportional to the distance, the differential term describes short range interaction which results from the interaction among the atoms on the surface plane. While $\beta_{e}$ and $\beta_{s}$ are taken to be zero namely the
discreteness effect is neglected the generalized 2D P-N equation can
be obtained \cite{Mryasov}.

\vskip 0.2in \noindent{\large \bf 3.2. The core properties of
dissociated dislocation with variational method}

The dislocation solution of 2D dislocation equation is unknown. In
this paper, the variational method is applied for the 2D dislocation
equation \cite{wang2002,wxz}. It can be straightforwardly verified
that the variational functional of the 2D dislocation equation takes
the following form
\begin{eqnarray}
\label{variational} J & = &-\frac{\beta_{e}}{4}\int_{-\infty}^{+\infty}\Big(\frac{du^{x}}{dx}\Big)^{2}dx-\frac{K_{e}}{4\pi}\int_{-\infty}^{+\infty}\int_{-\infty}^{+\infty}\frac{du^{x}}{dx}\frac{du^{x}}{dx}\Big|_{x=x^{'}}\ln\Big|\frac{x-x^{'}}{d}\Big|dxdx^{'} \nonumber \\
& & -\frac{\beta_{s}
}{4}\int_{-\infty}^{+\infty}\Big(\frac{du^{y}}{dx}\Big)^{2}dx-\frac{K_{s}}{4\pi}\int_{-\infty}^{+\infty}\int_{-\infty}^{+\infty}\frac{du^{y}}{dx}\frac{du^{y}}{dx}\Big|_{x=x^{'}}\ln\Big|\frac{x-x^{'}}{d}\Big|dxdx^{'} \nonumber \\
& & +\int_{-\infty}^{+\infty}\gamma (u^{x},u^{y})dx,
\end{eqnarray}
where $d$ is the distance of the nearest atoms in the misfit plane.

In order to obtain the structure of dissociated dislocation, we
represent each of the two partials by two closely spaced Peierls
dislocations (of arctan-type) that can assure the correct
asymptotic behavior. Thus, trial dislocation solution for
dissociated dislocation can be written as
\begin{eqnarray}\label{trial}
u^{x}&=&\frac{b}{2\pi} \Big\{ \arctan \Big(\frac{x+d_{eq}/2}{\zeta
_{e}})+\arctan \Big(\frac{x-d_{eq}/2}{\zeta _{e}}\Big)\Big\}
+\frac{b}{2}, \nonumber \\
u^{y}&=&\frac{\sqrt{3}b}{6\pi} \Big\{ \arctan
\Big(\frac{x+d_{eq}/2}{\zeta _{s}}\Big)-\arctan
\Big(\frac{x-d_{eq}/2}{\zeta _{s}}\Big) \Big\},
\end{eqnarray}
where $d_{eq}$ is the equilibrium separation of two partial
dislocations and $\pm d_{eq}/2$ are the positions of the two
partials, $\zeta _{e}$ and $\zeta _{s}$ represent the core width
of the edge and screw components respectively. The relation
between the edge and screw components of the relative displacement
determines the dissociation path in the glide plane. Generally,
the width of edge component $\zeta _{e}$ is not equal to that of
screw component $\zeta _{s}$ and the dissociation path deviations
from the crystallographic Burgers vector\cite{Mryasov}. Thus, the
dissociated dislocations cannot be dealt with 1D dislocation
equation.

Substituting the trial solution into the variational functional
Eq.(\ref{variational}), the total variational functional $J$ is as a
function of $d_{eq}$, $\zeta_{e}$ and $\zeta_{s}$. The separation
and the width of edge and screw components can be determined by
solving the following equations
\begin{equation}
\frac{\partial J}{\partial d_{eq}}=0,
\end{equation}
\begin{equation}
\frac{\partial J}{\partial \zeta_{e}}=0,
\end{equation}
\begin{equation}
\frac{\partial J}{\partial \zeta_{s}}=0.
\end{equation}

After a tedious although straightforward algebra one can obtain that
the variational parameters of the trial dislocation solutions
Eqs.(\ref{trial}) for all our calculated FCC metals. Table 3 gives
observed and calculated dissociated distances for the Al
edge-character partial dislocation, in comparison with the wide
range of previous results based on P-N model, atomistic simulations
and density-functional theory (DFT). The partial separation distance
in our results is $3.6b$ (where $b$ is length of the total Brugers'
vector), which is in good agreement with the experiment value $2.8b$
obtained from weak-beam transmission electron microscopy
\cite{Hollerbauer} and the simulated value $2.4b\sim3.4b$ from DFT
method in which the periodic supercell uses 552 atoms in a large
radius cylindrical slab surrounded by vacuum \cite{Woodward}, as
well as the value $3.2b$ obtained from the molecular dynamics
simulation with the glue potential \cite{Olmsted}. An earlier DFT
calculation, which employed the quasi-continuum (QC) DFT method,
where a small first principles cell containing the dislocation is
embedded in the strain field produced by atoms interacting through
an interatomic force law based on the embedded atom method (EAM),
gives a smaller dissociated distance $2.0b$ \cite{Lu2}. The
difference of the QC-DFT-EAM calculation most likely stems from the
small (84 atoms) supercell used for the first-principles
calculations. The P-N model, which is a semidiscrete model, is
essentially a continuum treatment. But the atomic-scale discreteness
of crystals really matters in the region of partial dislocation
core, so P-N models \cite{Lu2,SchoeckAl} are not well suited for
describing details of the strain field in partial core. Different
implementations of the P-N model predict different dissociated
distances, though both are based on first-principles calculations of
the GSFE surface. Several atomistic simulations for edge dislocation
in Al have obtained different results for the dissociated distance
between two partials, ranging from $1.7b\sim5.6b$
\cite{Olmsted,Srinivasan,Mishin}; this variation is possible the
result of different treatments for the boundary conditions
\cite{Woodward}. In our calculation, the GSFE surface is calculated
from first-principle calculations accurately and the 2D improved P-N
equation Eqs.(\ref{mpn}) fully includes the discreteness effect of
crystal, so our method is appropriate for dealing with the Shockley
partials which are the mixture dislocations including edge component
and screw component. The disregistry and the dislocation density
which is defined as $\rho=du/dx$ for Al, determined from our first
principles calculations combined with the improved P-N equation, are
shown in Fig. 3 (a) and (b), respectively. It is worth noting that
the maximum value of relative displacements along y-axis $u_{y}$ for
Al is $0.13b$, and is much smaller than the projection on the y-axis
of the Burgers' vector of the partial
$\frac{1}{6}\langle112\rangle=0.29b$ (the black dot in Fig.3(a)).
This means that the area within two partials is not a pure stacking
fault ribbon, where the core regions of the two partials overlap and
the displacement of y-component annihilate with each other due to
the opposite direction\cite{SchoeckAl,FangAl}. The structure of
dislocations in Al cannot be dealt with the 1D model due to the
strong overlap of two partials as illustrated in Fig.3(b). The two
partial dislocations can not be identified due to the large overlap
that is consistent with the experiment and the first principle
simulation \cite{Woodward}.

In this work, the core structures of edge dislocation in other FCC
metals Cu, Ir, Pd, and Pt have also been calculated by using first
principles calculations GSFE surfaces based on the improved P-N
equation. The core widths both of the edge and screw components and
the dissociated distances are listed in Table 4, and the disregistry
and the dislocation density for Cu, Ir, Pd, and Pt are shown in
Fig.4, Fig. 5, Fig.6, and Fig.7, respectively. We also listed the results obtained from the P-N model in which the second-derivative term is neglected in Table 4. One can see that the discrete effects always increase the widths both for edge and screw components.
Physically, the improved P-N equation Eq.(\ref{mpn}) is more functional to determine the structures of dislocation than the classical P-N equation \cite{wang2009}, but for the calculated FCC metals we regret not being able to find any experimental data which would clearly demonstrate that the results with second-derivative term are more superior than those without second-derivative term. We cannot be sure that which model is a better one at the present level of knowledge. However, we believe the new experimental results will clarify this question mark.

It has been known
that the deformation mechanism cannot be explained by the absolute
value of ISFE $\gamma_{\mathrm{ISF}}$ alone and the entire behavior
has to be understood in terms of the ratio $\gamma_{\mathrm
USF}/\gamma_{\mathrm ISF}$ along $\langle112\rangle$
\cite{Swygenhoven}, so we also list the ratios of
$\gamma_{\mathrm{USF}}/\gamma_{\mathrm{ISF}}$ in Table 4. When this
ratio is low, then the energy barrier that has to be overcome for
creating a trailing partial is very low and therefore it will be
possible to observe the full dislocation. Pt has the lowest ratio
$\gamma_{\mathrm{USF}}/\gamma_{\mathrm{ISF}}$ value $1.10$ in our
calculation, so the overlap of two partials is so strong that one
cannot distinguish them though dissociation behavior exists. In Fig
.7, we can see that the maximum value of $u_{y}$ for Pt is $0.11b$
and two-peak of the dislocation density for edge component
$\rho_{x}$ vanishes. But when this ratio is large, the energy
increase necessary for nucleating the trailing partial substantial,
which is in the case of Cu with the large ratio
$\gamma_{\mathrm{USF}}/\gamma_{\mathrm{ISF}}$ value 4.07. In Fig. 4,
we can indicate that the overlap of two partials is weak in Cu.  It
has verified in simulations (Ref\cite{schiotz}) extended partial
dislocations in Cu have been observed as the predominant deformation
mechanism at nanocrystalline grains. It is worth noting that though
the dissociated distance $5.5b$ for Pt is larger than that $5.4b$
for Pd, the core width of Pt is larger than that of Pd. This
explains why Fig. 6 for Pd shows two-peak of $\rho_{x}$ but Fig .7
for Pt does not. Our detailed calculations demonstrate that whether
full dislocations or Shockley partials will easily be observed in
FCC lattice must be understood in terms of the ratio
$\gamma_{\mathrm{USF}}/\gamma_{\mathrm{ISF}}$.

\section{conclusion}

In conclusion, we have performed first-principles calculations to
obtain the GSFE surface for the  $\{111\}$ glide plane in FCC
metals Cu, Al, Ir, Pd, and Pt. The accuracy of the method have been
tested by calculating values for various stacking fault energies
which favorably compare with the previous theoretical and
experimental results. From these calculations, we extract the core
properties for the edge dislocation of dissociation into partials,
using the improved P-N equation which includes the discreteness
effect of crystal. The Ritz variational method is presented to solve
the dislocation equation and the trial solution is constituted by
two arctan-type functions which represents the dislocation
dissociating into two partials. The core structures, including the
core widths both of the edge and screw components, dissociation
behavior for edge dislocations have been investigated.
 The dissociated distance between two partials for Al in our calculation
agrees well with the values obtained from the numeric simulation with DFT
and MD methods, as well as experiment. Our detailed calculations
demonstrate that whether full dislocations or Shockley partials will
easily be observed in FCC lattice must be understood in terms of the
ratio $\gamma_{\mathrm{USF}}/\gamma_{\mathrm{ISF}}$, and show that
it is preferred to create partial dislocation in Cu, and to be
easily observed full dislocation in Al, Ir, Pd, and especially Pt

 \vskip 2in

\footnotesize

\def\refname{

\newpage
 \vskip 0.4 in  {\footnotesize {\noindent \bf Table  1} The calculated equilibrium lattice parameters $a$ [{\AA}] and elastic constants [GPa], comparison
  with the experimental and previous calculated results.

\begin{tabular}{ccccccc}
\multicolumn{7}{l}{}\\\hline
\ \  & a & $c_{11}$ &$c_{12}$ & $c_{44}$ & $\mu$ & $\nu$ \\
\hline
Al &4.05, $4.05^{a}$   &112.3, $114.3^{b}$ &60.7, $61.3^{b}$& 30.2, $31.6^{b}$&25.8, $26.3^{b}$&0.350, $0.349^{b}$ \\
Cu &3.64, $3.62^{a}$   &175.3, $176.2^{b}$ &124.4, $124.9^{b}$& 80.9, $81.8^{b}$&25.5, $25.7^{b}$&0.415, $0.415^{b}$ \\
Ir &3.88, $3.84^{a}$   &579.6, $582.3^{b}$ &240.8, $241.3^{b}$& 261.2, $262.0^{b}$& 167.9, $170.5^{b}$& 0.294, $0.293^{b}$ \\
Pd &3.96, $3.89^{a}$   &232.1, $234.1^{b}$ &175.2, $176.1^{b}$&70.5, $71.2^{b}$&28.5, $29.0^{b}$& 0.430, $0.420^{b}$ \\
Pt &3.99, $3.92^{a}$   &356.8, $358.0^{b}$ &252.2, $253.6^{b}$&76.3, $77.4^{b}$&  52.3, $52.2^{b}$&0.414, $0.415^{b}$ \\
 \hline
\end{tabular}}

\begin{tabular}{cccccccc}
 \leftline {$^{a}$Reference\cite{Handbook1}} \\
 \leftline {$^{b}$Reference\cite{Handbook2}}\\
\end{tabular}

\vskip 0.4in
{\footnotesize {\noindent \bf Table  2} The various
stacking-fault energies on
 (111) plane for FCC metals. The experimental and the other
 calculated results are also listed. All values are in units
 of $\mathrm{mJ/m^{2}}$.

\begin{tabular}{cccccc}
\multicolumn{6}{l}{}\\\hline
\ \  & Cu & Al & Ir & Pd & Pt  \\
$\gamma_{1}$ & -160& -158&  -615&-167 &-171\\
$\gamma_{2}$ & -260& -213&  -928&-243 &-188\\
$\gamma_{\mathrm{ISF}}$ & 43& 158&  359&122 &282\\
$\gamma_{\mathrm{ISF}}$$^{a}$  &75&146&41&244&356 \\
$\gamma_{\mathrm{ISF}}$$^{b}$ &43&150&- &101&111 \\
$\gamma_{\mathrm{ISF}}$$^{c}$ &51 &153&  -& 186&- \\
$\gamma_{\mathrm{ISF}}$$^{d}$ &40, 45, 169&166&300&180&322 \\
 $\gamma_{\mathrm{USF}}$ along $\langle112\rangle$ &175&225& 753&215  & 311 \\
 $\gamma_{\mathrm{USF}}$ along $\langle110\rangle$ &638& 633&  2462&  668&  685 \\
 \hline
\end{tabular}

\begin{tabular}{cccccccc}
 \leftline {$^{a}$Reference\cite{Zhang1}, by EAM calculations}; \\
 \leftline {$^{b}$Reference\cite{Wei}, by MEAM calculations;}\\
 \leftline {$^{c}$Reference\cite{Hartford}, by first-principles calculations;}\\
 \leftline {$^{d}$Reference\cite{Hirth,Grujicic,Thorntor}, by experiments.}\\
      \end{tabular}}

\vskip 0.4 in  {\footnotesize {\noindent \bf Table  3} Shockley
partial dissociated distance $d_{eq}$ for $1/2\langle110\rangle$
edge dislocations in Al. Our calculated result is compared with the
experimental results ,  the previous theoretical values including the
Peierls-Nabarro methods and the numerical simulations. All data  are
in units of the length of Burgers vector $b=\sqrt{2}/2a$.

\begin{tabular}{ccccccccc}
\multicolumn{9}{l}{}\\\hline
 This work  & $^{\mathrm{a}}$Ref\cite{Hollerbauer} & $^{\mathrm{b}}$Ref\cite{Woodward} & $^{\mathrm{b}}$Ref\cite{Lu1} & $^{\mathrm{c}}$Ref\cite{Lu2}
  &$^{\mathrm{c}}$Ref\cite{SchoeckAl} & $^{\mathrm{d}}$Ref\cite{Wang} &$^{\mathrm{d}}$Ref\cite{Olmsted} & $^{\mathrm{d}}$Ref\cite{Srinivasan}\\
 \hline
 3.6  & 2.8 & $2.5\sim3.4$ & 2.0  &1.2 & 2.6 & 3.2 & 5.1 & 5.6, $1.7\sim5.2$ \\
\hline

\end{tabular}

\begin{tabular}{cccccccc}
 \leftline {$^{a}$ experiment from weak-beam transmission electron microscopy (WB-TEM)}; \\
 \leftline {$^{b}$ obtained from density-functional theory (DFT);}\\
 \leftline {$^{c}$ obtained from Peierls-Nabarro model;}\\
 \leftline {$^{d}$ obtained from atomistic simulations}\\
      \end{tabular}}

\vskip 0.4in {\footnotesize {\noindent \bf Table  4} The predicted
core structures, including core widths both the edge and screw
components, dissociated distance in FCC metals Cu, Al, Ir, Pd, and Pt are listed. For completeness, we also list the results obtained from the P-N model in which the second-derivative term is neglected.  For
discussion, we  give the ratio
$\gamma_{\mathrm{USF}}/\gamma_{\mathrm{ISF}}$ values along the
$\langle112\rangle$ direction. The ratio of
$\gamma_{\mathrm{USF}}/\gamma_{\mathrm{ISF}}$ is dimensionless, and
$\zeta_{e}$, $\zeta_{s}$ and $d_{eq}$ are in units of the length of
Burgers vector $b=\sqrt{2}/2a$.

\begin{tabular}{cccccc}
\multicolumn{6}{l}{}\\\hline
\ \  & Cu & Al & Ir & Pd & Pt  \\
$\zeta_{e}$ & 2.6$^{a}$, 2.1$^{b}$& 2.5$^{a}$, 2.0$^{b}$&3.5$^{a}$, 3.1$^{b}$&3.1$^{a}$, 2.4$^{b}$&4.8$^{a}$, 4.0$^{b}$\\
$\zeta_{s}$ & 1.9$^{a}$, 1.5$^{b}$& 2.1$^{a}$, 1.6$^{b}$&2.9$^{a}$, 2.5$^{b}$&2.4$^{a}$, 1.8$^{b}$&3.9$^{a}$, 3.0$^{b}$\\
$d_{eq}$    & 5.5$^{a}$, 3.7$^{b}$& 3.6$^{a}$, 2.6$^{b}$&5.6$^{a}$, 4.6$^{b}$&5.4$^{a}$, 3.5$^{b}$&5.5$^{a}$, 4.0$^{b}$\\
$\gamma_{\mathrm{USF}}/\gamma_{\mathrm{ISF}}$& 4.07&1.42&2.10& 1.76&1.10 \\
 \hline
\end{tabular}

\begin{tabular}{cccccccc}
 \leftline {$^{a}$ obtained from improved P-N equation (with second-derivative term)}; \\
 \leftline {$^{b}$ obtained from P-N equation (without second-derivative term).}\\
 \end{tabular}
\newpage

\vskip 0.4in
\begin{center}
\scalebox{0.5}[0.5]{\includegraphics{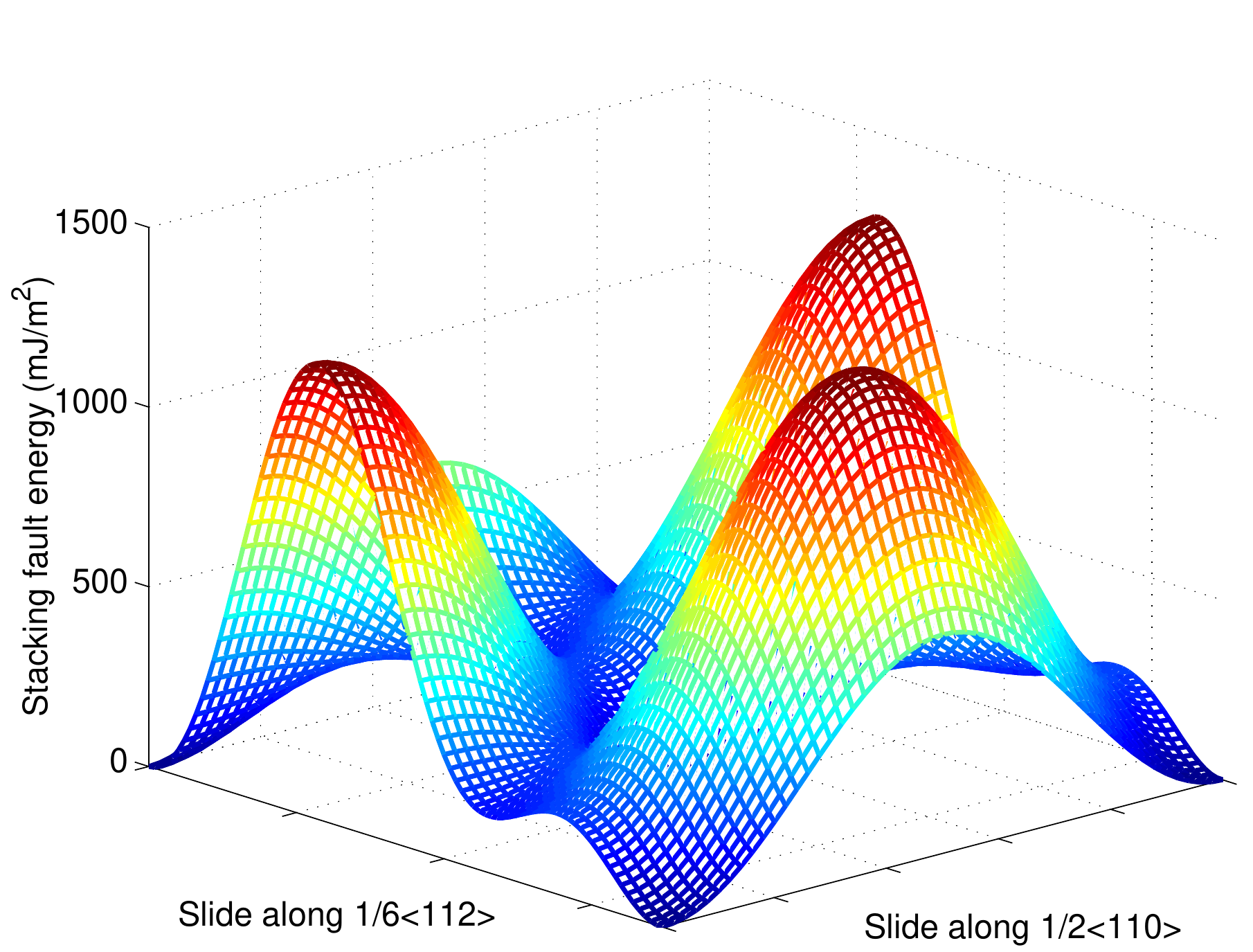}}
\end{center}
{\footnotesize{{\bf Fig. 1} The GSFE surface for displacements along
a (111) plane in Al (the corners of the plane and its center
correspond to the configurations of ideal Al lattice). The first
energy maximum encountered along the $\langle112\rangle$ direction
is the USFE; the first energy minimum value at
1/6$\langle112\rangle$ corresponds to the ISF configuration. The
projection of the GSFE surface along the
1/2$\langle110\rangle$ is symmetric with respect to the slip
displacement of $\sqrt{2}a/4$.}}

\vskip 0.4in
\begin{center}
\scalebox{0.5}[0.5]{\includegraphics{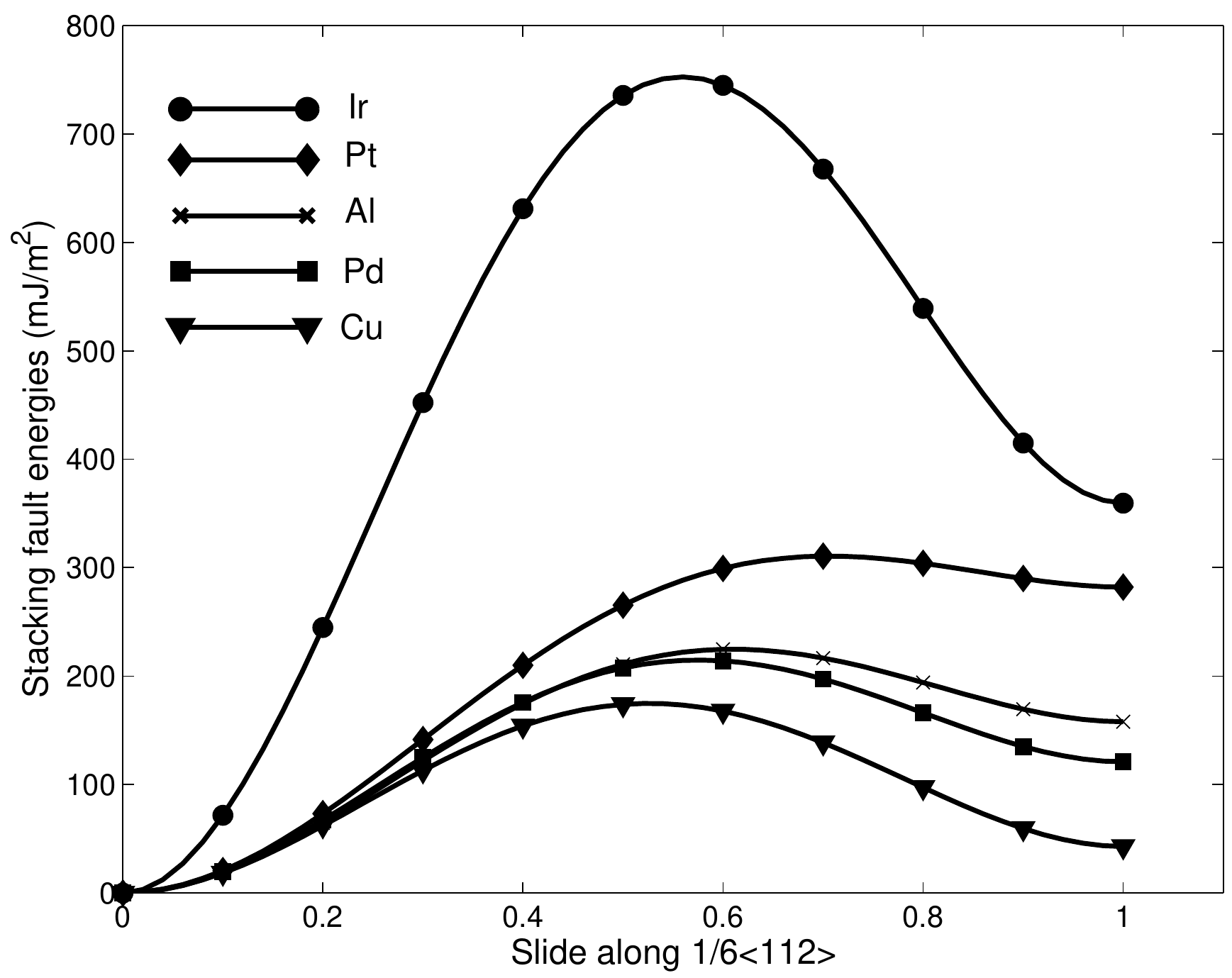}}
\end{center}
{\footnotesize{{\bf Fig. 2} Projection of the $\gamma$-surface in
the [11\={2}] directions on the (111) plane. There have been termed
the intrinsic stacking fault (ISF) energies corresponding to the
final positions and the unstable stacking fault (USF) energies
corresponding to the maximum of the curves.}}

\vskip 0.4in
\begin{center}
\scalebox{0.5}[0.6]{\includegraphics{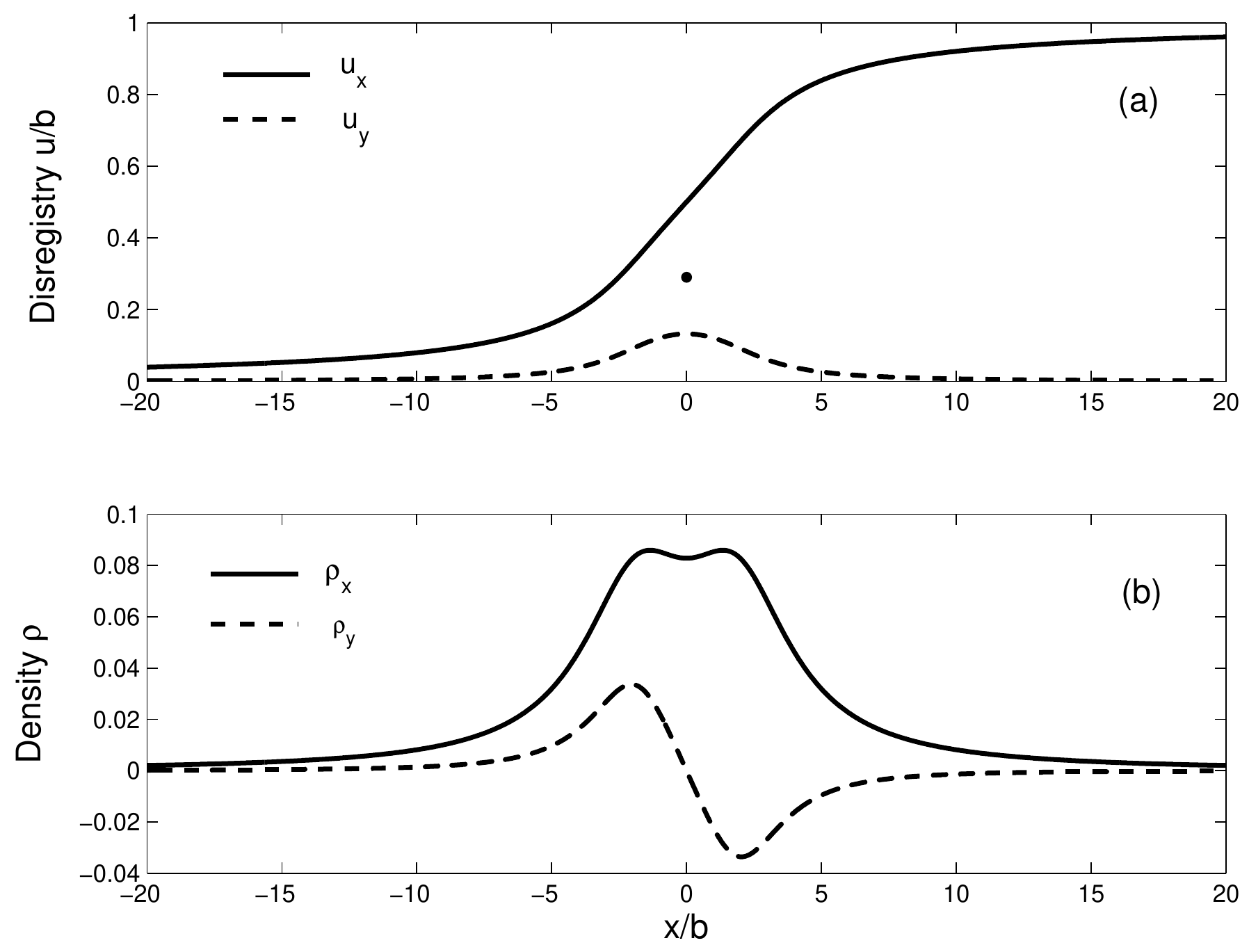}}
\end{center}
{\footnotesize{{\bf Fig. 3.} Disregistry profile u(x) (a) and
dislocation density (b) for the edge dislocation in Al are obtained
from our first-principles calculations combined with the improved
P-N equation. The solid line and the dashed line represent the edge
and screw components of displacements, respectively. The black dot
marks the displacement of the screw component of the
crystallographic Shockley partial dislocation, which is about
$0.29b$. Here the maximum value of relative displacements of screw
components of the partial dislocation is 0.13b.}}

\vskip 0.4in
\begin{center}
\scalebox{0.5}[0.6]{\includegraphics{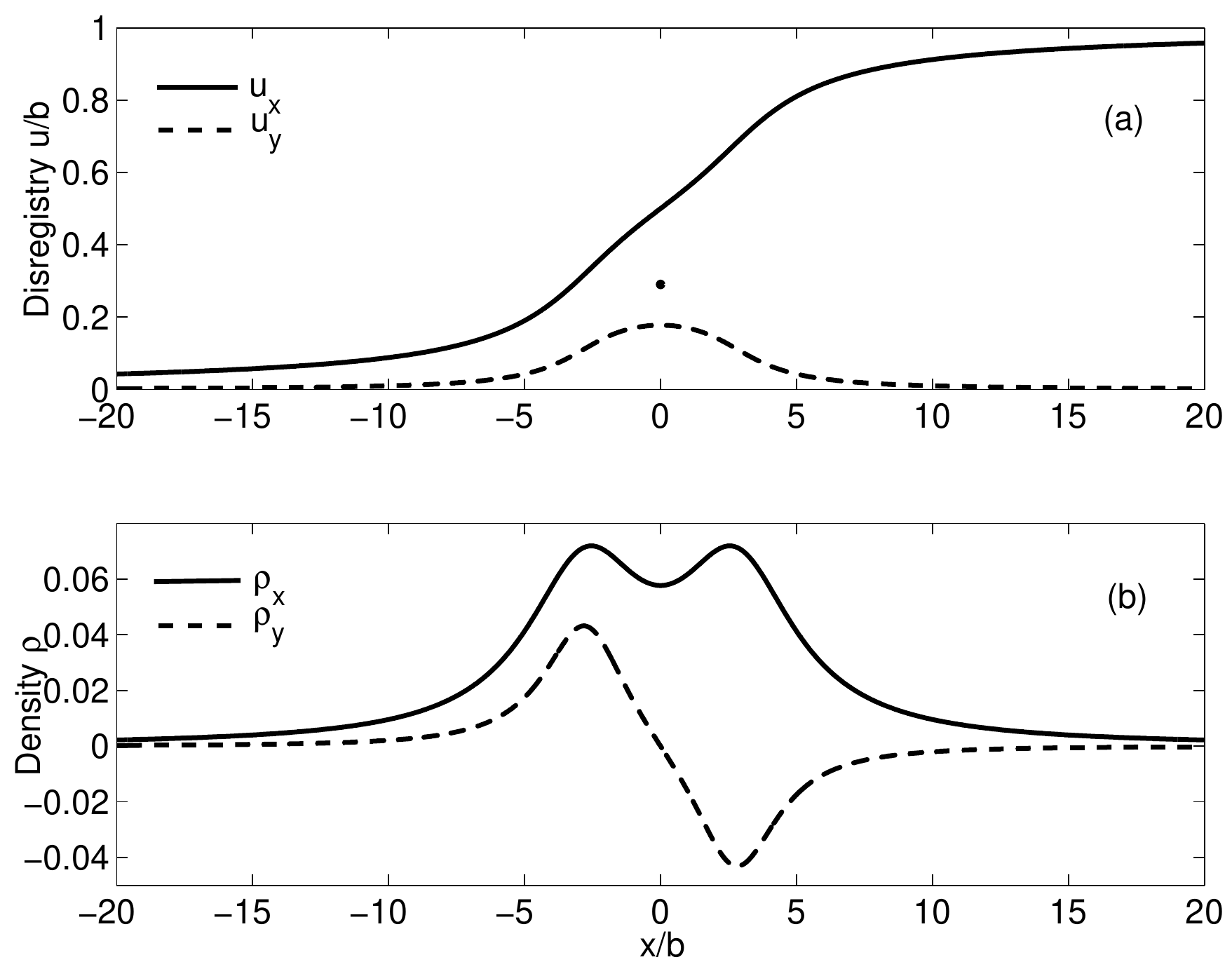}}
\end{center}
{\centerline{\footnotesize{{\bf Fig. 4.} Same as in Fig. 3, but for
Cu. Here the maximum value of relative displacements of screw
components of the partial dislocation is 0.20b.}}}

\vskip 0.4in
\begin{center}
\scalebox{0.5}[0.6]{\includegraphics{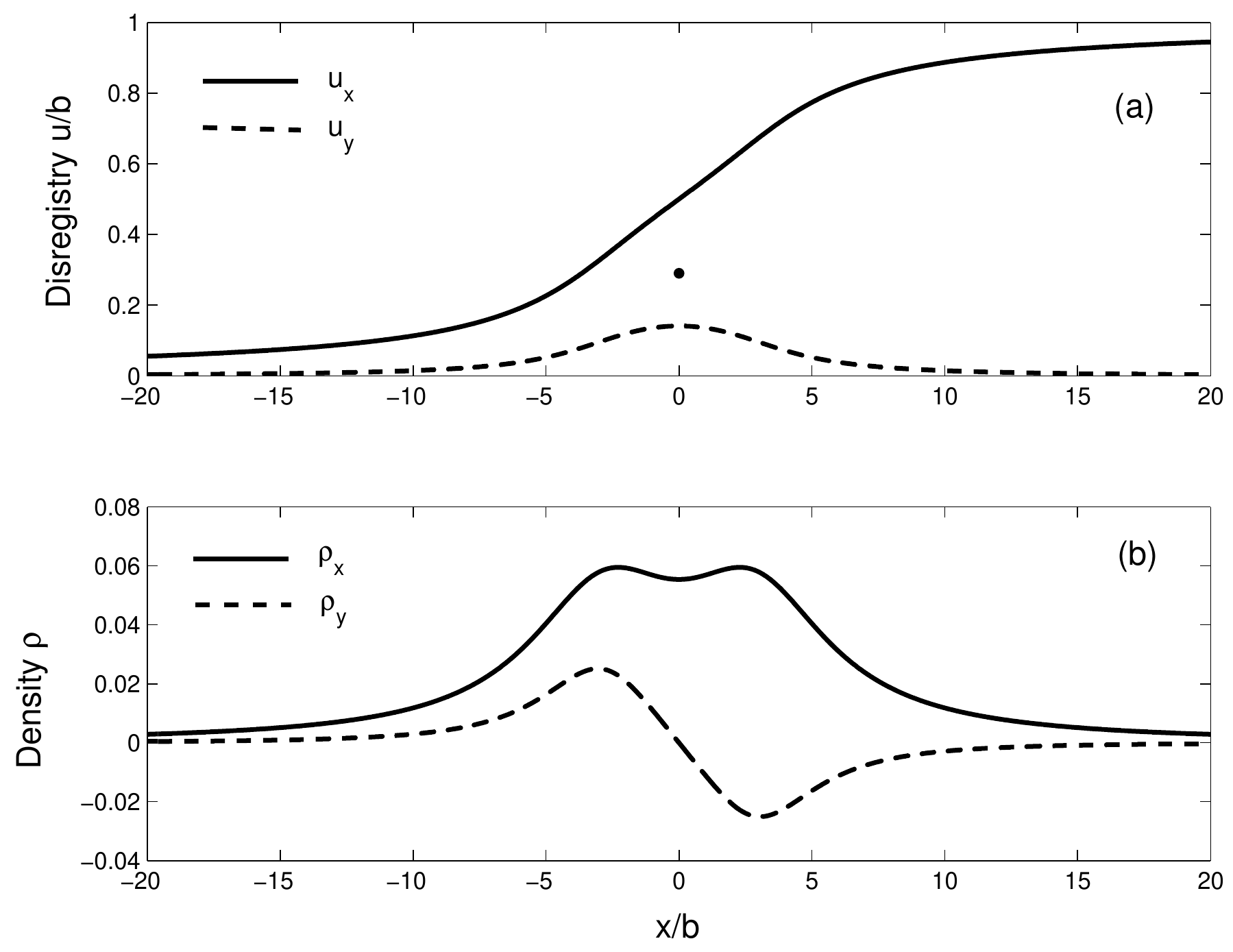}}
\end{center}
{\centerline{\footnotesize{{\bf Fig. 5.} Same as in Fig. 3, but for
Ir. Here the maximum value of relative displacements of screw
components of the partial dislocation is 0.15b.}}}

\vskip 0.4in
\begin{center}
\scalebox{0.5}[0.6]{\includegraphics{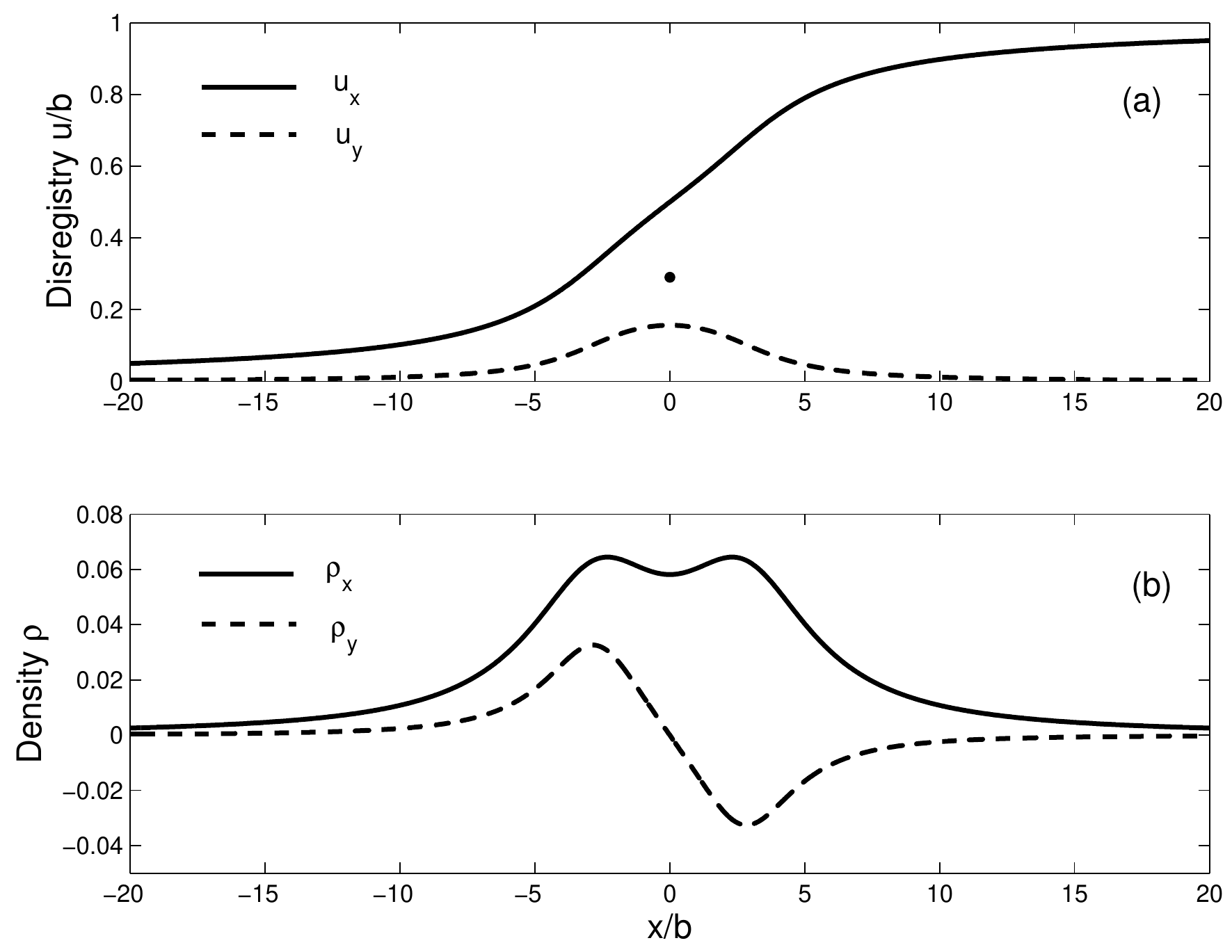}}
\end{center}
{\centerline{\footnotesize{{\bf Fig. 6.} Same as in Fig. 3, but for
Pd. Here the maximum value of relative displacements of screw
components of the partial dislocation is 0.15b.}}}

\vskip 0.4in
\begin{center}
\scalebox{0.5}[0.6]{\includegraphics{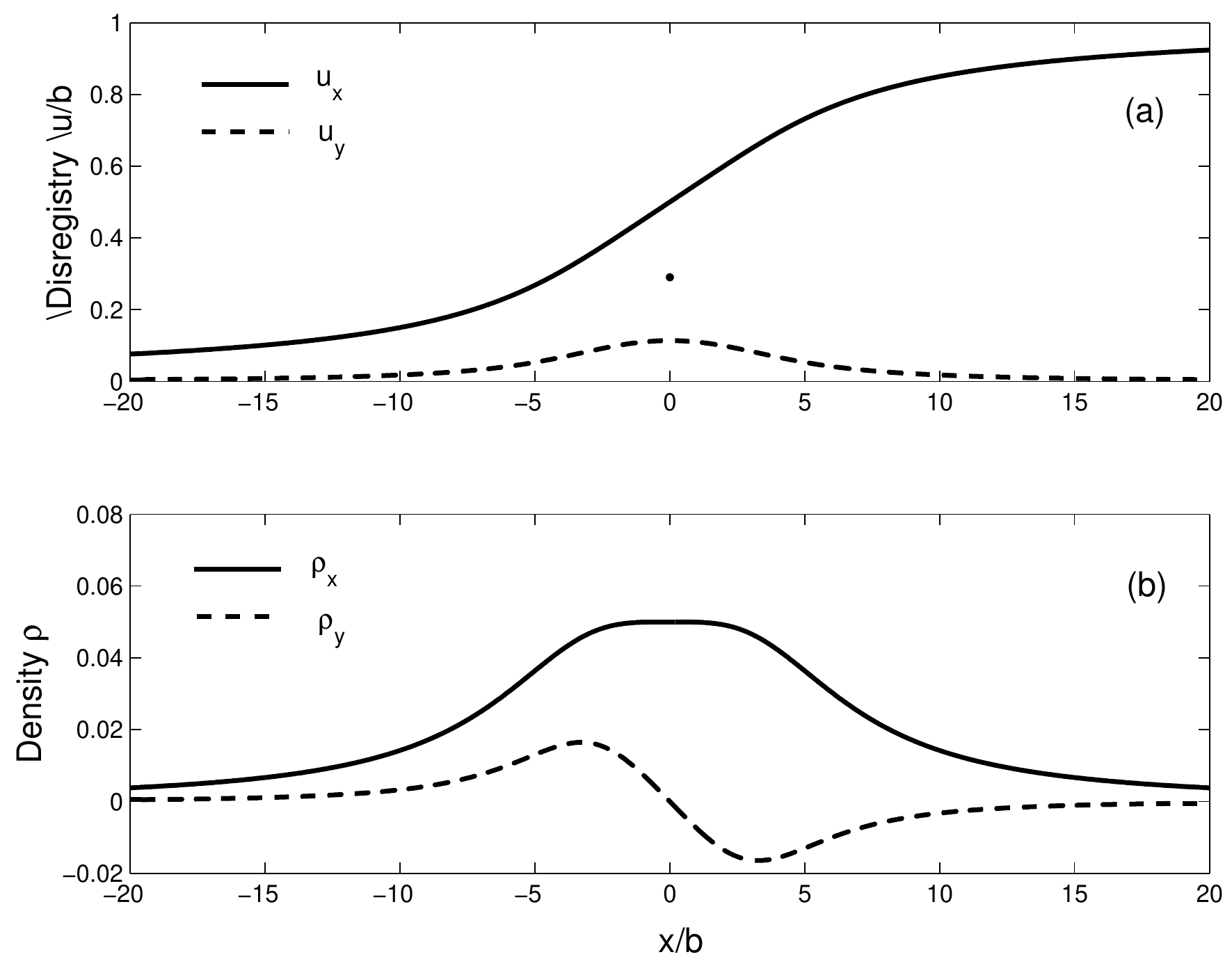}}
\end{center}
{\centerline{\footnotesize{{\bf Fig. 7.} Same as in Fig. 3, but for
Pt. Here the maximum value of relative displacements of screw
components of the partial dislocation is 0.11b.}}}

\end{document}